\documentclass[12pt]{article}

\usepackage{english,a4}

\usepackage{epsfig}
\usepackage{dcolumn}
\usepackage{amsmath}
\usepackage{cite}
\usepackage{changebar}

\makeatletter \@addtoreset{equation}{section}
\makeatother 
\renewcommand{\thefootnote}{\alph{footnote}}

\newcommand{\scr}[1] {\mbox{\scriptsize #1}}

\newcommand{\DT} {\ensuremath{\Delta(T)\times T^{4}}}

\newcommand{\tc} {\ensuremath{T_{\scr{c}}}}

\begin{document}
\thispagestyle{empty}
%
%\hbox{}
% \mbox{} \hspace{1.0cm}
         \today 
\mbox{} \hfill BI-TP 99/20\hspace{1.0cm}\\
\mbox{} \hfill ESI-733-1999\hspace{1.0cm}\\
\begin{center}
\vspace*{1.0cm}
\renewcommand{\thefootnote}{\fnsymbol{footnote}}
{\LARGE\bf ANOMALOUS CURRENTS AND~\\
  GLUON CONDENSATES IN QCD~\\ AT FINITE TEMPERATURE
\footnote{Presented at the~{\bf Workshop on Quantization, 
Generalized BRS Cohomology and Anomalies}~\\ at the Erwin
Schr\"odinger International Institute for Mathematical
Physics in Vienna, Austria on~\\ 2. October 1998} \\}

\vspace*{1.0cm}
{\large David E. Miller$^{1,2}$
\\}
\vspace*{1.0cm}

${}^1$ {Fakult\"at f\"ur Physik, Universit\"at Bielefeld, Postfach 100131,\\ 
D-33501 Bielefeld, Germany}\\

${}^2$ {Department of Physics, Pennsylvania State University,
Hazleton Campus,\\
Hazleton, Pennsylvania 18201, USA (permanent address)}\\
\vspace*{2cm}
{\large \bf Abstract \\}
~\\
After a short description of the currents coming from the known
conservation laws in classical physics, we look at some further 
cases which arise after quantization in relation to quantum
chromodynamics (QCD) at finite temperature. In these cases,
however, some basic changes appear with the anomalies. 
First we go into the relationship between the trace of the
energy momentum tensor and the gluon condensates at finite
temperatures. Using the recent numerical data from the simulations 
of lattice gauge theory we present the computational evaluations 
for the gluon condensates at finite temperatures. Thereafter we 
discuss the effects of chiral symmetry breaking and its restoration 
at finite temperature through the chiral phase transition. 
In this context we investigate the properties of the gluon condensate 
in the presence of massive dynamical quarks using the numerical data. 
Finally we put together these results with a discussion of the 
various anomalous currents and their relationship to our findings here. 

\end{center}




\section{\bf Introduction}
\noindent    
It is well known that the Maxwell Equations are invariant under the
ten generators of the Lorentz Group. This property may be expressed
by means of the conservation laws for the terms corresponding to the
energy-momentum (4 terms) and the angular momentum (6 terms). This
was the understanding already at the founding of the special theory of 
relativity~\cite{Ein}. Not long thereafter Bateman~\cite{Bate} 
pointed out that electromagnetism in vacuuo had a much greater 
invariance under the fifteen generator conformal group.
Some of the consequences of this were later pointed out in a  
work of Bessel-Hagen~\cite{BesHag}, which came out of an
investigation of the conservation laws due to Noether~\cite{Noet}
in both mechanics and electrodynamics with special attention to the 
corresponding currents. This work~\cite{BesHag} made a careful analysis of the
conformal group and its realization in the additional currents known as 
the special conformal and dilatation currents, which appear in addition 
to those of the conserved energy, momentum and angular momentum. Furthermore,
it was then~\cite{BesHag} realized that the special conformal (4 terms)
and the dilatation (1 term) currents are not conserved when massive particles 
are present. This fact may be expressed by the presence of a trace in the
energy momentum tensor. However, Bessel-Hagen then proceeded to set this 
trace to zero in order to get the sought after conservation equations.
This procedure is clearly valid in classical electrodynamics when no
massive particles are present. The obtained conservation laws represent
all the currents invariant under the full conformal group in Minkowski
space-time. This is historically the classically expected situation for a
pure gauge invariant field theory.
~\\
\indent
    About one half a century ago Steinberger as well as Fukuda and Miyamoto
studied the electromagnetic decay of the neutral pion into two photons
$\pi^{0}~\rightarrow~2\gamma$. They regarded this process as going over to a
virtual pair of proton and antiproton written in the form of the now famous 
triangle diagram where $\pi^{0}~\rightarrow~{{p}{\bar{p}}}~\rightarrow~2\gamma$
. This brought about a number of studies of related ideas especially that 
of Schwinger who pointed out that the conservation of the axial current 
$J^{\mu}_{5}$ given by $\bar{\psi}{\gamma^{\mu}}{\gamma_{5}}\psi$
in QED is not upheld when the current is properly regularized.
This fact is stated by the divergence of this axial current in the equation
\begin{equation}
\partial_{\mu}J^{\mu}_{5}~=~J_{5}~+
{{{e^2}k}\over{8\pi^2}}{{\bar{F}^{\mu\nu}}F_{\mu\nu}},
\label{eq:chiranom}
\end{equation} 
where $J_{5}$ is $2imP$ with $P$ as $\bar{\psi}{\gamma_{5}}\psi$
the expected pseudoscalar part of the  divergence relating
to the particle mass $m$, k is a constant containing some numerical factors 
like $\hbar$ and $F^{\mu\nu}$ and ${\bar{F}^{\mu\nu}}$ are
the electromagnetic field strength tensor and its dual.
This result was later rediscovered in 1969 independenly by Adler in spinor 
electrodynamics and Bell and Jackiw \cite{AdBeJa} for the $\sigma$ model.
Their results were very similar to the above form in~\eqref{eq:chiranom}. 
A discussion of the axial current anomaly is found in the
literature~\cite{BertKugo,TJZW}. We will look into the properties of the
axial currents in relation to the breaking of chiral symmetry in QCD
and its restoration at very high temperatures.
~\\
\indent
     Now we want to look at these currents to discuss them in relation to the 
physics arising in the presence of strong interactions. We shall assume a
local conservation of energy and momentum under the strong interaction.
 During the course of this work we shall see how the unconserved currents
in QCD at finite temperature come to provide particular differential forms. 
First we look at the dilatation current $D^{\mu}$ and the four special 
conformal currents $K^{\mu\alpha}$. 
We begin with the equation for $D^{\mu}(x)$ which is just the
product $x_{\alpha}T^{\mu \alpha}$ of the displacement four vector
$x^{\mu}$ and the energy momentum tensor $T^{\mu\nu}$ whose
four-divergence gives after renormalization just the trace of the energy-
momentum tensor $T^{\mu}_{\mu}$ as follows:
\begin{equation}
\partial_{\mu} D^{\mu}~=~T^{\mu}_{\mu}.
\label{eq:dilcurr}
\end{equation} 
\noindent
A similar equation can be written for the divergence of the special conformal
currents
\begin{equation}
\partial_{\mu}K^{\mu\alpha}~=~2{x^{\alpha}}T^{\mu}_{\mu},
\label{eq:specconf}
\end{equation} 
\noindent
where the special conformal currents $K^{\mu\alpha}$ are given by
\begin{equation}
K^{\mu\alpha}~=~(2{x^{\alpha}}{x_{\nu}}~-~g^{\alpha}_{\nu})T^{\mu\nu}.
\label{eq:specconfcurr}
\end{equation} 
\noindent
Prior to around 1970 it was generally supposed that the finiteness of
$T^{\mu}_{\mu}$ related~{\it{only}} to known masses. The renormalization of
the nonabelian field theories and the study of the renormalization group
equations (Callan-Symanzik equations) brought new attention to the problem.
Now this brings us to the point of actually considering what is new in QCD 
at finite temperatures. Furthermore, we also want to know why all the 
conservation laws of classical electromagnetism do not fulfill our
expectations. The simple answer to these questions lie in the process of 
renormalization of the quantum field theory, which acts as a scale setter.
We shall also discuss the chiral anomaly both for its historical role
as well as its role in the presence of dynamical quarks.
~\\
\indent 
     In the next section we shall first discuss the consequences for the 
gluon condensate at finite temperature. Thereafter we use the results of 
numerical simulations for the pure gauge theories~\cite{Eng4,Boyd}.
The essential relationship for these calculations is the trace anomaly 
which arises directly from the scale variance of QCD. It relates the trace 
of the energy momentum tensor to the square of the gluon field strenths
through the renormalization group beta function. Here we shall
expand upon the approach investigated in~\cite{Mill}, for which the
consequences of the new finite temperature lattice data for $SU(N_{c})$
gauge theory for the gluon condensate~\cite{BoMi} have been presented.
After this we shall look into some of the properties of the chiral condensate
at finite temperature in relation to the axial anomaly in QCD, which relates
to the presence of the quark condensates at finite temperatures. At this
point we will consider from the numerical results the properties of the 
gluon condensate in the presence of dynamical quarks. Then we mention some 
of the properties of the anomalous currents at finite temperature in terms 
of the related differential forms to which we ascribe a certain physical 
meaning. Finally we conclude this work with a brief discussion of a
three dynamical quark model



\section{\bf The Trace Anomaly at finite Temperature}
\noindent
The study of the relationship between the trace of the energy momentum
tensor and the gluon condensate has been carried out at finite temperatures 
by Leutwyler~\cite{Leut} in relation to the problems of deconfinement 
and chiral symmetry. He starts with a detailed discussion of the 
trace anomaly based on the interaction between Goldstone bosons 
in chiral perturbation theory. Central to his discussion is the 
role of the energy momentum tensor, whose trace is directly related
to the gluon field strength. It is important to note that the
energy momentum tensor $T^{\mu\nu}(T)$ can be separated into the zero 
temperature or confined part, $T^{\mu\nu}_{0}$, and the finite temperature
contribution $\Theta^{\mu\nu}(T)$ as follows:
\begin{equation}
  \label{eq:emtensor}
  T^{\mu\nu}(T) = T^{\mu\nu}_{0} + \Theta^{\mu\nu}(T) .
\end{equation}
\noindent
The zero temperature part, $T^{\mu\nu}_{0}$, has the standard problems with
infinities of any ground state. It has been discussed by 
Shifman, Vainshtein and Zakharov \cite{SVZ1} in relation to the 
nonperturbative effects in QCD and the operator product expansion. 
The finite temperature part, which is zero at $T=0$, 
is free of such problems. We shall see in the next section how
the diagonal elements of $\Theta^{\mu\nu}(T)$ are calculated in a
straightforward way on the lattice. The trace  $\Theta^{\mu}_{\mu}(T)$ 
at finite temperatures in four dimensions is connected to  
the thermodynamical contribution to the energy density 
$\epsilon(T)$ and pressure $p(T)$ for relativistic fields as well as 
in relativistic hydrodynamics~\cite{LaLi}
\begin{equation}
  \Theta^{\mu}_{\mu}(T) = \epsilon(T) - 3p(T) .
  \label{eq:eps-ideal}
\end{equation}
\noindent
This quantity actually provides a form of the equation of state!
The gluon field strength tensor is denoted by
$G^{\mu\nu}_a$, where $a$ is the color index for $SU(N)$.
The basic equation for the relationship between the gluon condensate
and the trace of the energy momentum tensor at finite temperature was
written down \cite{Leut} using the trace anomaly in the form as Leutwyler's
equation
\begin{equation}
\langle G^2 \rangle_{T} = \langle G^2 \rangle_0~-~
\langle {\Theta}^{\mu}_{\mu} \rangle_{T} , 
\label{eq:condef}
\end{equation}
\noindent
where the gluon field strength squared summed over the colors is
\begin{equation}
G^{2}~=~{{-\beta(g)}\over{2g^3}} G^{{\mu}{\nu}}_{a}G_{{\mu}{\nu}}^{a} , 
\end{equation}
\noindent
for which the brackets with the subscript $T$ mean thermal average.
The renormalization group beta function $\beta(g)$ in terms of the
coupling may be written as
\begin{equation}
\beta(g)~=~\mu{dg \over{d\mu}}
         =~-{1 \over{48\pi^{2}}}(11N_{c}~-~2N_{f})g^{3}~+~O(g^{5}) .
\label{eq:betafun}
\end{equation}
\noindent
The quantity $\beta(g)$ has the effect of summing the loop contributions
arising in the renormalization of the vertices. In the third order $O(g^3)$
the structure starts with the triangle diagram for the vertex correction, 
which for the pure gluon case has just the two contributions at this order.
However, the case for light quarks additional terms are included 
in the renormalization process involving the quark triangle as well as the
ghost triangle not to mention the further explicit effects of the mass
renormalization~\cite{Muta}.

\indent
Leutwyler has calculated for two massless quarks using the low temperature
chiral perturbation expansion the trace of the energy momentum tensor at 
finite temperature in the following form:
\begin{equation}
\langle \Theta^{\mu}_{\mu} \rangle_{T}~=~{\pi^{2} \over {270}}
{T^{8} \over {F^{4}_{\pi}}}{\left\{ln{{\Lambda_{p}}\over{T}}\right\}}
~+~O(T^{10}),
\end{equation}
\noindent
where the logarithmic scale factor $\Lambda_{p}$ is about $0.275~GeV$ and
the pion decay constant $F_{\pi}$ has the value of $0.093~GeV$. The value
of the gluon condensate for the vacuum $\langle G^{2} \rangle_{0}$ was
taken to be about $ 2~GeV/fm^{3}$, which is consistent with the previously
calculated values \cite{SVZ1}. The results sketched by Leutwyler at Quark
Matter'96 in Heidelberg \cite{Leut} show a long flat region for
$\langle G^{2} \rangle_{T}$ as a function of the temperature until it
arrives at values of at least $0.1~GeV$ where it begins to show a falloff
from the vacuum value proportional to the power $T^{8}$.



\section{\bf Lattice Data for the Gluon Condensate for Pure $SU(N_c)$} 

In this section we want to describe the lattice computation at finite 
temperature in some detail. As usual for statistical physics we start with 
a partition function ${\cal{Z}}(T,V)$ for a given temperature T and spatial 
volume $V$. From this we may define the free energy density as follows:
\begin{equation}
          f~=~-{T\over{V}}{\ln{\cal{Z}}(T,V)}.
\end{equation}
The volume $V$ is determined by the lattice size $N_{\sigma}a$, where $a$ is
the lattice spacing and $N_{\sigma}$ is the number of steps in the given
spatial direction. The inverse of the temperature $T$ is determined by
$N_{\tau}$ is the number of steps in the (imaginary)temporal direction. 
Thus the simulation is done in a four dimensional Euclidean space 
with given lattice sizes ${N_{\sigma}^3}\times{N_{\tau}}$, which gives 
the volume $V$ as $(N_{\sigma}a)^3$ and the inverse temperature $T^{-1}$ 
as $N_{\tau}a$ for the four dimensional Euclidean volume. In an $SU(N_{c})$ 
gauge theory the lattice spacing $a$ is a function of the bare gauge
coupling $\beta$ defined by $2N_{c}/g^2$, where g is the bare $SU(N_{c})$
coupling. Thereby this function fixes both the temperature and the volume 
at a given coupling. Now we let $P_{\sigma,\tau}$ the expectation value of,
respectively space-space and space-time plaquettes, whereby
\begin{equation}
    P_{\sigma,\tau}~=~1~-~{1\over{N}}
     {Re{\langle{Tr(U_{1}U_{2}U^{\dagger}_{3}U^{\dagger}_{4})}}}
\end{equation}
\noindent
for the usual Wilson action~\cite{Boyd}. These plaquettes may be generalized
to the improved actions on anisotropic lattices~\cite{EngKarSch} for SU(2) 
and SU(3). For the symmetric Wilson action we define the parts $S_{0}$ as
$6P_{0}$ on the symmetric lattice $N^{4}_{\sigma}$ and $S_{T}$ as
$3(P_{\sigma}+P_{\tau})$ on the asymmetric lattice 
${N^{3}_{\sigma}}\times{N_{\tau}}$. We now proceed to compute the free energy
density by integrating these expectation values as

\begin{equation}
 {f(\beta)\over{T^4}}~=~-{N^{4}_{\tau}}{\int\limits_{\beta_{0}}^{\beta}}
                        {d{\beta'}[S_{0}~-~S_{T}]},
\end{equation}
where the lower bound $\beta_{0}$ relates to the constant of normalization.
At this point we should add that the free energy density is a fundamental
thermodynamical quantity from which all other thermodynamical quantities
can be gotten. Also it is very important in relation to the phase structure
of the system in that the determination of the transitions for their order
and critical properties as well as the stability of the individual phases 
are best studied.
~\\
\indent
Next we define lattice beta function in terms of the lattice spacing $a$ and
the coupling $g$ as
\begin{equation}
  {\tilde{\beta}(g)}~=~-2{N}{a\frac{dg^{-2}}{da}}.
\end{equation}
\noindent
The dimensionless interaction measure $\Delta(T)$~\cite{Engels}  
is then given by
\begin{equation}
  \Delta(T) = {N^{4}_{\tau}}~{\tilde{\beta}(g)}~{\left[S_{0}~-~S_{T}\right]}.
\end{equation}
\noindent
The crucial part of these recent calculations is the use of the full
lattice beta function, $\tilde{\beta}(g)$ in obtaining
the lattice spacing $a$, or scale of the simulation,
from the coupling $g^{2}$. Without this accurate information
on the temperature scale in lattice units it would not be possible
to make any claims about the behavior of the gluon condensate.
The interaction measure is the thermal ensemble expectation value given by
 $(\epsilon - 3p)/T^4$. Thus because of equation~\eqref{eq:eps-ideal} above 
the trace of the temperature dependent part of the energy momentum tensor
here denoted as $\Theta^{\mu}_{\mu}(T)$ is equal to the expectation value
of $\Delta(T)$ multiplied by a factor of $T^4$,
which may be calculated~\cite{Mill,BoMi} as a function of the temperature as
\begin{equation}
  \Theta^{\mu}_{\mu}(T)~=~\DT.
\label{eq:trace}
\end{equation}
\noindent
There are no other contributions to the trace for the pure gauge fields on 
the lattice. The heat conductivity is zero. Since there are no non-zero 
conserved quantum numbers and, as well, no velocity gradient in the lattice 
computations, hence no contributions from the viscosity terms appear.
For a scale invariant system, such as a gas of free massless particles,
the trace of the energy momentum tensor, equation~\eqref{eq:trace}, is zero.
A system that is scale variant, perhaps from a particle mass, 
has a finite trace, with the value of the 
trace measuring the magnitude of scale breaking. At zero temperature it 
has been well understood from Shifman et al.~\cite{SVZ1}
how in the QCD vacuum the trace of the energy momentum tensor
relates to the gluon field strength squared, $G^{2}_0$.
Since the scale breaking in QCD occurs explicitly at all orders in a loop
expansion, the thermal average of the trace of the energy momentum tensor
should not go to zero above the deconfinement transition.
So a finite temperature gluon condensate $G^{2}(T)$
related to the degree of scale breaking at all temperatures, can be defined
to be equal to the trace. We have used~\cite{BoMi} the lattice simulations
~\cite{Eng4,Boyd} in order to get the temperature dependent part 
of the trace and, thereby, the value of the condensate at finite temperature.
The trace of the energy momentum tensor as a function of the temperature
is shown in Figure 1. We notice that for $T~<~\tc$ it remains constant
at zero. However, above $\tc$  in both cases there is a rapid rise in
${\Theta}^{\mu}_{\mu}(T)$. 
Accordingly, the vacuum gluon condensate $G^{2}_{0}$ becomes just the usually
assumed value $0.012~GeV^4$  for both cases~\cite{SVZ1}.
It is clear that newer estimates~\cite{Nari} attribute a considerably higher
value to the vacuum gluon condensate of about $0.0226~GeV^4$, which clearly
increases our value for the low temperature phase, but does not in any way
alter the conclusions for the pure gauge system in as far as the disappearance
of the condensate is concerned.
By taking the published data~\cite{Eng4,Boyd} for $\Delta(T)$, and using
Leutwyler's  equation~\eqref{eq:condef} together with the equation 
for the trace at finite temperature ~\eqref{eq:trace} we have obtained 
the gluon condensate $G^{2}(T)$ at finite temperature as shown in Figure 1.
\begin{figure}[b]
   \begin{minipage}{75mm}
  \begin{center}
    \leavevmode
    \epsfig{file=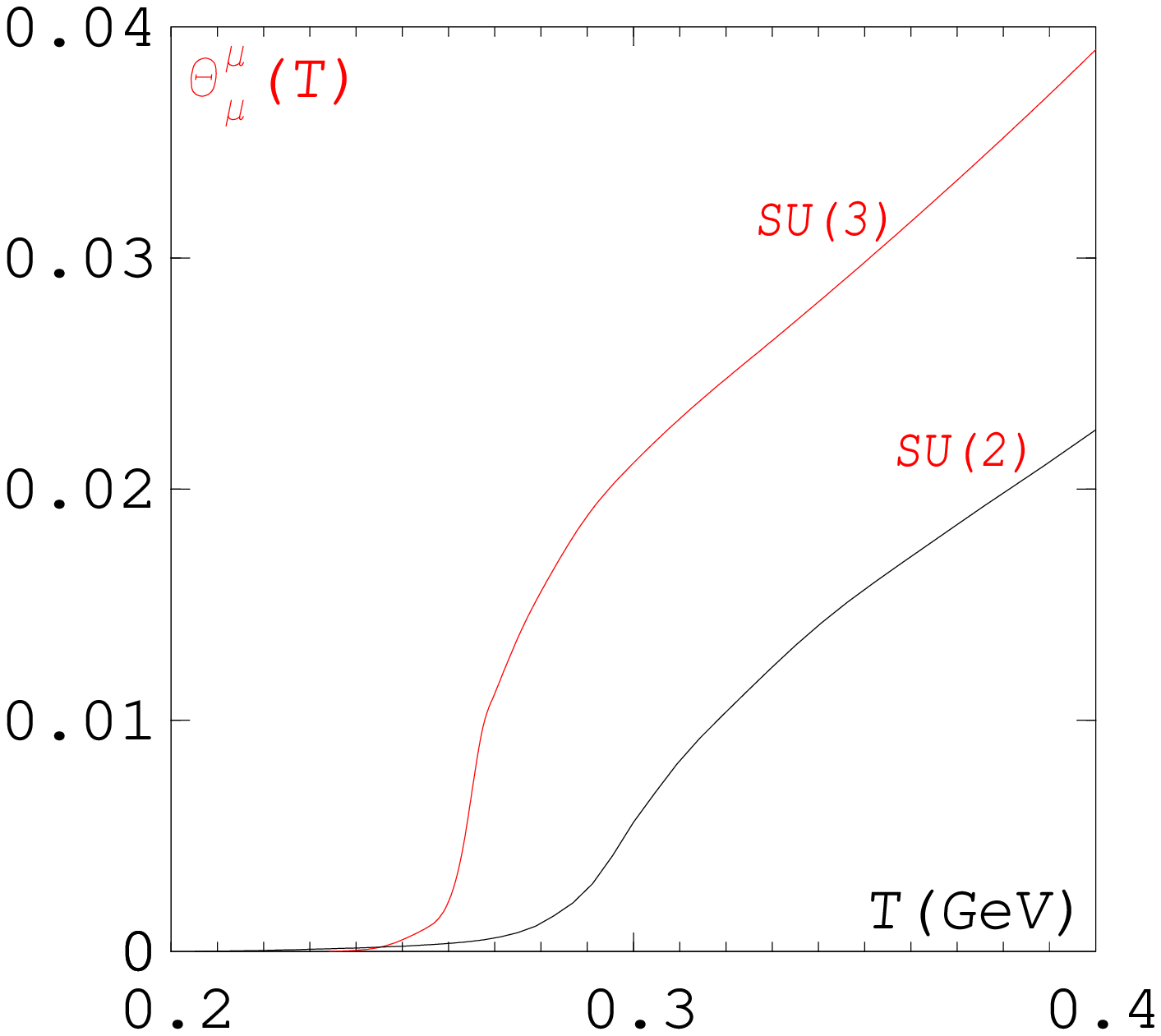
        ,height=85mm
        ,bbllx=85,bblly=200,bburx=550,bbury=650}
   \end{center}
   \end{minipage}
\hfill   
\begin{minipage}{75mm}
  \begin{center}
    \leavevmode
      \epsfig{file=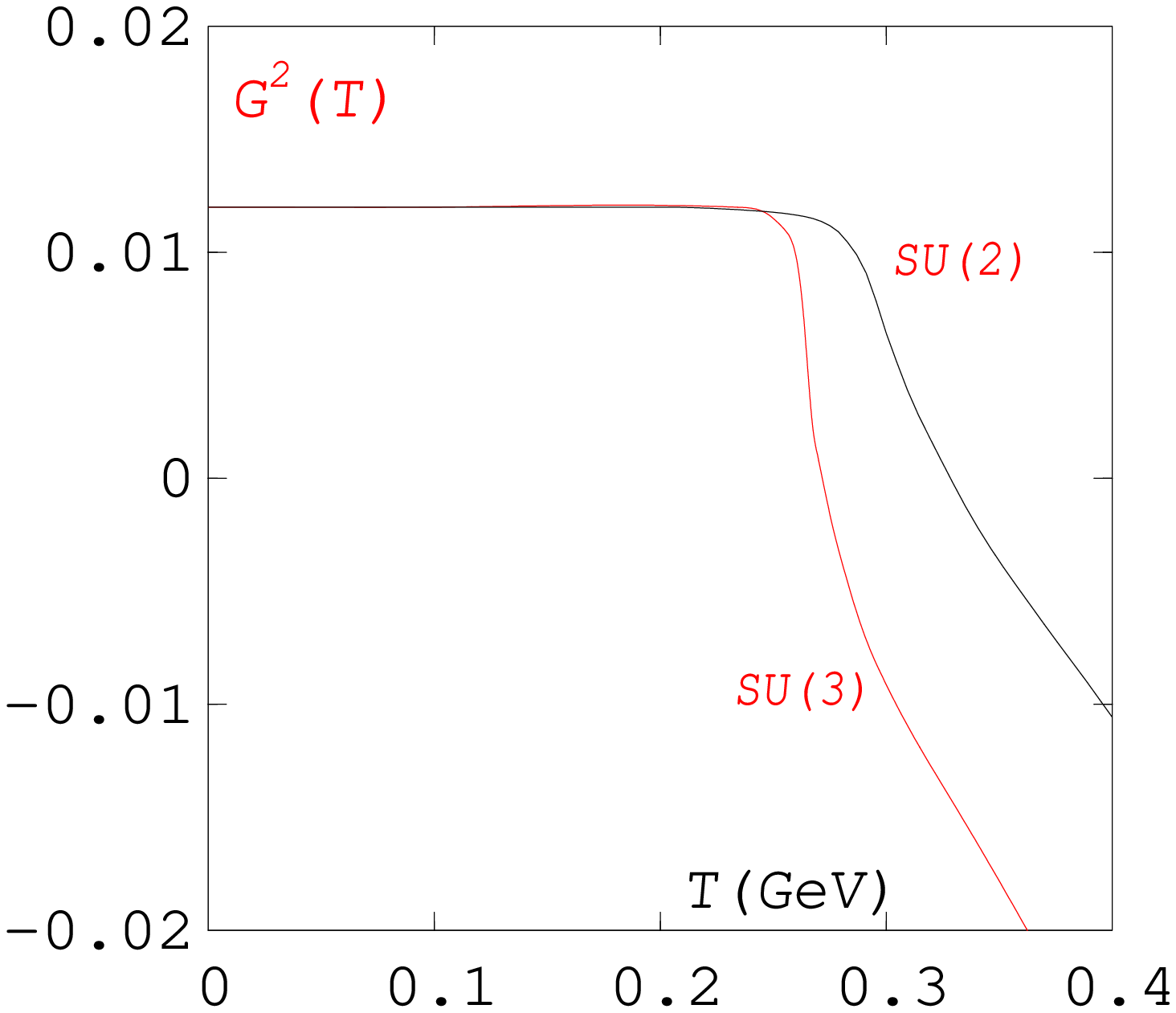,bbllx=85,bblly=200,bburx=550,bbury=650,
          height=85mm}
     \end{center}
   \end{minipage}
   \caption{      \label{fig:cond3}
   The left plots show ${\Theta}^{\mu}_{\mu}(T)$ for the lattice gauge
   theories  $SU(2)$ and $SU(3)$ as indicated. The right plots show the
   corresponding gluon condensates. The values for the critical temperature
   $\tc = 0.290, 0.264$~GeV for $SU(2)$ and $SU(3)$, respectively. Both
   ordinates are in $GeV^4$.
          }
\end{figure}
In the left part of Figure 1 we compare the growth of $SU(2)$ and $SU(3)$
for the finite temperature part of the trace of the energy momentum tensor
${\Theta}^{\mu}_{\mu}(T)$.
We note the continued growth of the trace with increasing  temperature for the
pure gauge theories. It was already contrasted~\cite{BoMi} over a much 
larger temperature range the rapidity of growth in comparison where also 
properties of the critical behavior were included. The difference in the 
change of the thermal properties of the gluon condensate are then apparent 
with the assumed same vacuum structure, which is shown in the right part of 
Figure 1. These results of pure lattice gauge theory is completely consistent 
with the previous statement of Leutwyler~\cite{Leut}~"the gluon condensate
does not disappear but becomes negative and large." For the pure gauge theories
there appears no reason for stopping the melting process as the temperature
increases since gluons can be created without the limitations of a set scale
other than that which comes out of the renormalization process at any given
temperature.



\section{The Chiral Condensate at Finite Temperature}

In the presence of dynamical quarks another symmetry becomes important--
the chiral symmetry. When the quarks have masses, this symmetry is 
automatically broken. The chiral symmetry is a property of the two different
representations of $SL(2,\bf{C})$ denoted by $\bf{2}$ and $\bf{2^{*}}$
arising for the Dirac spinors. 
It is the presence of the quarks' mass terms in the 
Dirac equation that formally breaks the chiral symmetry. This comes 
formally out of the nonconservation of the axial current $j^{\mu}_{5}$
as discussed above in equation~\eqref{eq:chiranom} relating to the triangle 
diagrams, such that the chiral anomaly for QCD takes the form
\begin{equation}
\partial_{\mu}j^{\mu}_{5}~=~j_{5}~+
{{k_{1}}\over{8\pi^2}}{{\bar{G}^{\mu\nu}_{a}}G^{a}_{\mu\nu}},
\label{eq:chiranomqcd}
\end{equation} 
\noindent
where $k_{1}$ is a constant. This situation has important implications in
the case for finite temperatures where for $T$ sufficiently high the chiral
symmetry is restored in the small mass limit. We shall discuss the implications
of this both from the theoretical side and the numerical side where a finite
small mass is present.
~\\
\indent
    We now look at the chiral condensate at finite temperatures using chiral
perturbation theory. The low temperature expansion for two massless
quarks can be written~\cite{Leut} in the following form:
\begin{equation}
\label{eq:chirpert}
{{\langle \bar{\psi}_q{\psi}_q \rangle_{T}}\over{\langle \bar{\psi}_q{\psi}_q
  \rangle}_{0}}~=~1~-~{1\over{8}}{T^{2}\over{F^{2}_{\pi}}}~-~
{1\over{384}}{T^{4}\over{F^{4}_{\pi}}}~-~{1\over{288}}
{T^{6} \over {F^{6}_{\pi}}}{\left\{ln{{\Lambda_{q}}\over{T}}\right\}}
~+~O(T^{8})~+~O(\exp{-M\over{T}}),
\end{equation}
\noindent
where $F_{\pi}$ is the above mentioned pion decay constant and the scale
$\Lambda_{q}$ is taken as approxamately $0.470GeV$. Leutwyler has shown at
Quark Matter '96~\cite{Leut} that this expansion up to three loops remains 
very good at least to $0.100GeV$. Thus at low temperatures the probability
of finding any given excited mass state is related to the exponentially
small correction, which then has, indeed, a very little value. As the 
temperature grows the number of particle states begins to grow exponentially
as would be indicated by the Hagedorn spectrum~\cite{Haged},which leads to a
problem with this series at high temperatures. However, at low temperatures 
the excited states may be regarded as a dilute gas of free particles since 
the chiral symmetry supresses the interactions by a power of $T$ of this gas 
of excited states with the primary pionic component.
~\\
\indent
     Upon approaching the chiral symmetry restoration temperature $T_{\chi}$
the picture changes drastically.
At this point the ratio $T/F_{\pi}$ is considerably greater than unity. It is
here where one expects the chiral condensate to be very small or to have 
totally to have vanished. This effect has been studied recently numerically
~\cite{Laer} for two light flavors at finite temperature on the lattice.
The results of this simulation is shown for ${
{\langle \bar{\psi}_q{\psi}_q \rangle_{T}}/{\langle \bar{\psi}_q{\psi}_q
  \rangle}_{0}}$, which we simply write as $\langle \bar{\psi} {\psi}\rangle$.
We show this quark condensate ratio as a function of the coupling $\beta$
for the range where the chiral symmetry is largely restored~\cite{Laer}.
The left figure shows this ratio for two light quarks with a mass in lattice
units of $0.02$ on a lattice of size~$16^{3}\times 4$. The right figure
shows different mass values from left to right of $0.02$, $0.0375$ and $0.075$
on  $8^{3}\times 4$, $12^{3}\times 4$ and $16^{3}\times 4$ lattices. 
We should notice how the larger mass values slow the restoration down, 
which corresponds to moving the transition~$T_{\chi}$ to higher temperatures
or even eliminating it altogether as indicated by the flatness of the curves. 
~\\
\indent
The main quantities which were analyzed here were the various susceptibilities:
~\\
\noindent
1. The Polyakov loop susceptibility;
\begin{equation}
{{\chi}_{L}}~=~N^{3}_{\sigma}{\left[ \langle L^2 \rangle~
-~{\langle L \rangle}^2 \right]},
\label{eq:poloopsusc}
\end{equation} 
~\\
\noindent
2. The magnetic or chiral susceptibility;
\begin{equation}
{{\chi}_{m}}~=~{T\over V}~{\sum\limits_{i=1}^{N_{f}}}
{{\partial^2}\over{\partial m^{2}_{i}}}{\ln{\cal{Z}(T,V)}},
\label{eq:magsusc}
\end{equation}
~\\ 
\noindent
3. The thermal susceptibility;
\begin{equation}
{{\chi}_{\theta}}~=~-{T\over V}~{\sum\limits_{i=1}^{N_{f}}}
{{\partial^2}\over{\partial m_{i} \partial(1/T)}}{\ln{\cal{Z}(T,V)}}.
\label{eq:thermsusc}
\end{equation}
~\\ 
\noindent
One compares the critical properties of ${{\chi}_{L}}$,
${{\chi}_{m}}$ and ${{\chi}_{\theta}}$ in order 
to establish the value of $T_{\chi}$ and its
critical properties in the chiral limit where $m_{i} \rightarrow 0$.
However, in numerical simulations $m_{i}$  must be taken to be finite-- 
this means that one must use various different small values of $m_{i}$ on
different sized lattices $N^{3}_{\sigma}\times N_{\tau}$.
The procedure uses the lattice data to find the values around the peak of
the susceptibility ${{\chi}_{m}}$ at $T_{\chi}$ for the smallest masses,
with which one can determine the critical structure. A careful determination
of the topological susceptibility relating to the chiral current correlations
can be related to the square of the topological charge 
$Q^{2}_{T}$~\cite{HanLeut} in the chiral limit, such that
\begin{equation}
{N_{f}\over{m}}\langle Q^{2}_{T} \rangle~=~
      V{\langle \bar{\psi} {\psi}\rangle}_{m \rightarrow 0}.
\label{eq:topcond}
\end{equation}
Thus from the susceptibility one can arrive at the quark condensate
$\langle \bar{\psi} {\psi}\rangle$. However, in this computation it is a
major problem to properly set the temperature scale for small lattices with
finite masses. The plots in Figure 2 are made with the coupling $\beta$
which may be compared with pure $SU(3)$ on one side and the two flavor
dynamical quark simulations on the other~\cite{MILC97}. In the case of
pure $SU(3)$ the critical coupling $\beta_{c}$ for a ${16^3}\times 4$
lattice has the value~\cite{Boyd} of about 5.70, which is considerably 
larger than the values of $\beta$ shown in Figure 2. However, for the
two flavored dynamical quarks~\cite{MILC97} the value of ${\beta}_{c}$
is around 5.40, which is still somewhat above the values shown in this
figure.
\begin{figure}[b]
   \begin{minipage}{75mm}
  \begin{center}
    \leavevmode
      \epsfig{file=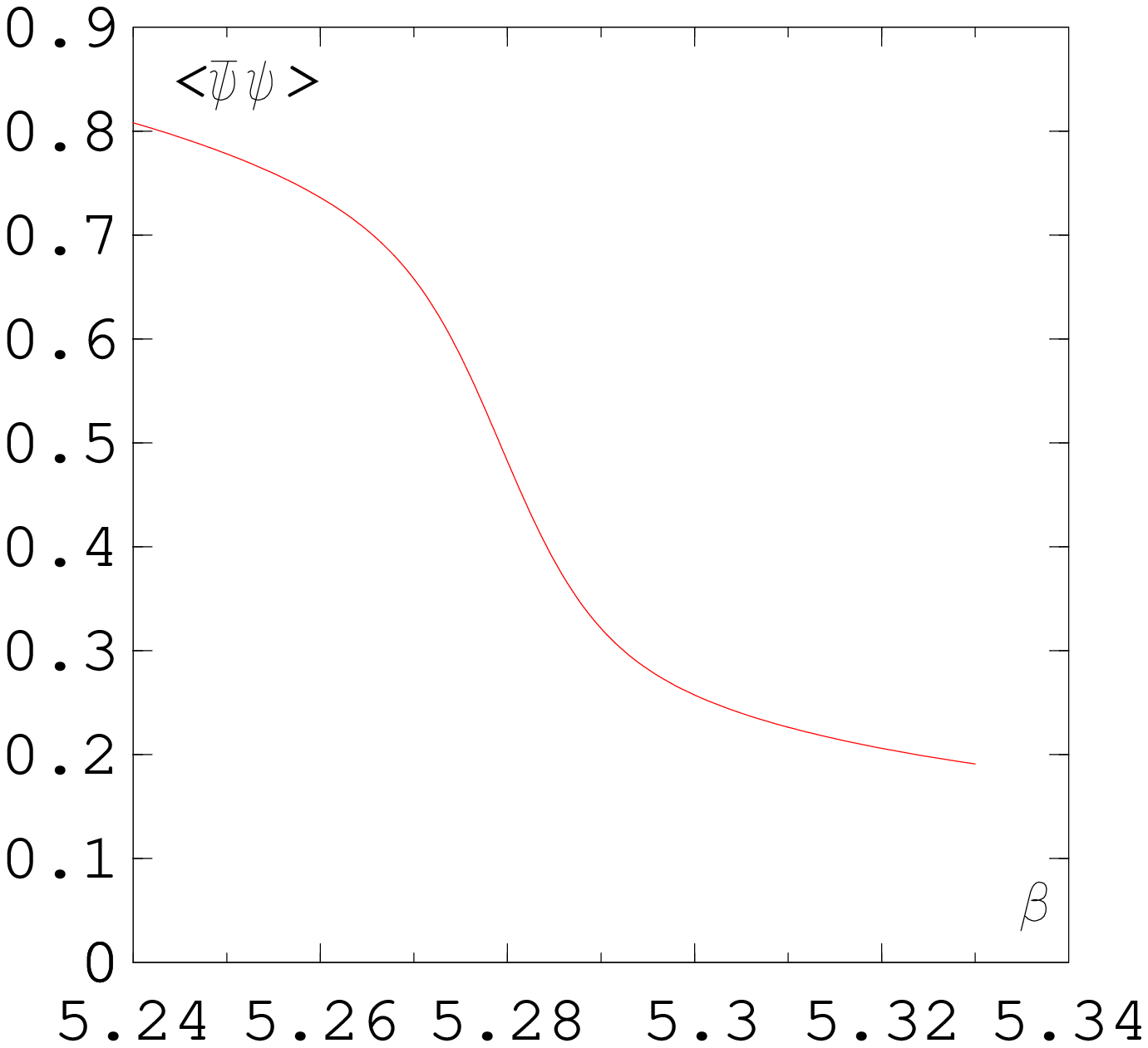,bbllx=85,bblly=200,bburx=550,bbury=650,
          height=85mm}
   \end{center}
   \end{minipage}
\hfill   
\begin{minipage}{75mm}
  \begin{center}
    \leavevmode
      \epsfig{file=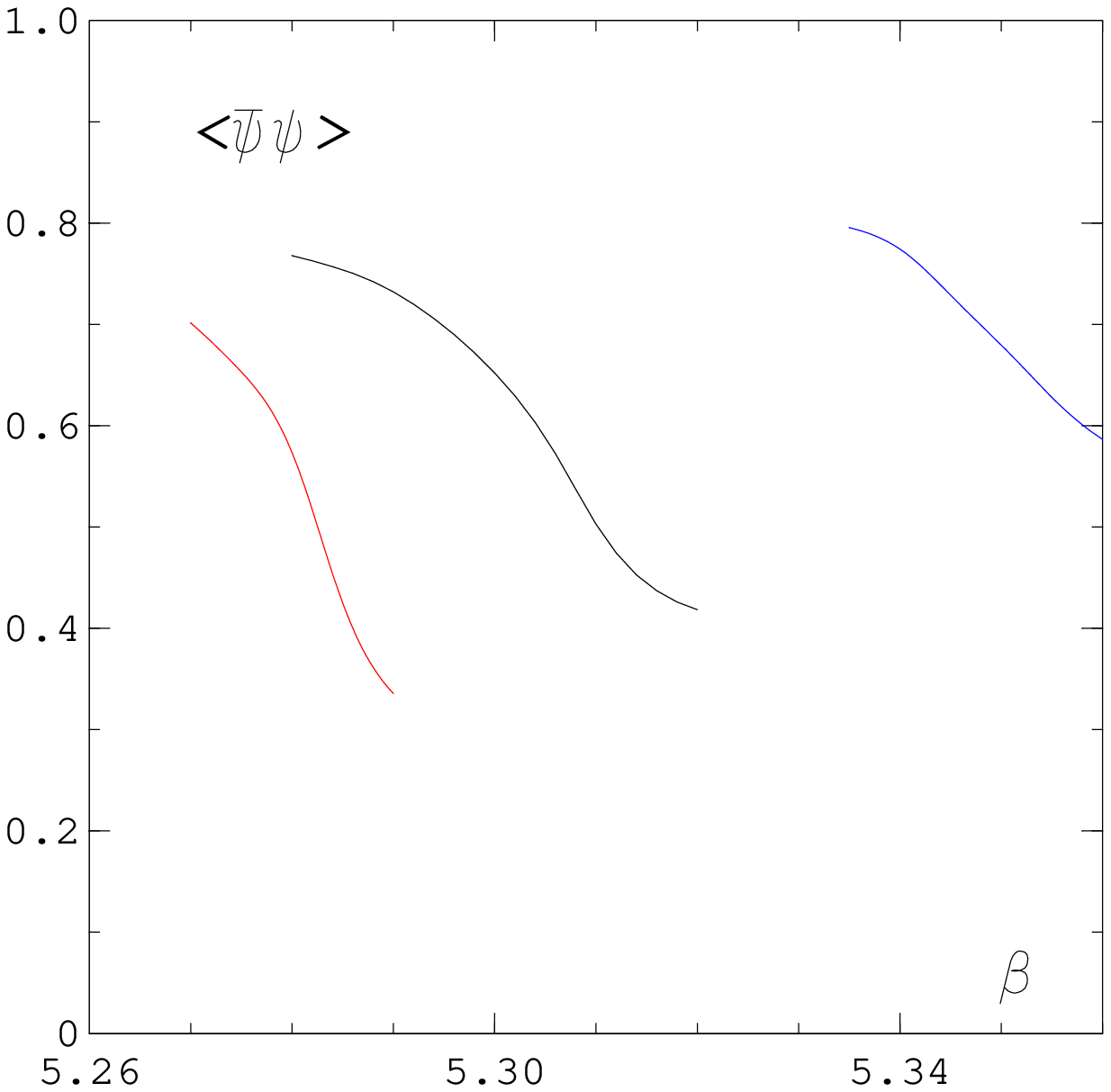,bbllx=85,bblly=200,bburx=550,bbury=650,
          height=85mm}
     \end{center}
   \end{minipage}
   \caption{      \label{fig:chir}
The left figure shows $<\bar\psi\psi >$ as a function of the coupling $\beta$
for the quark mass in lattice units $ma$= 0.02, which is normalized to the
vacuum value of the chiral condensate. The right figure shows the same 
quantity with the different quark masses 0.02, 0.0375 and 0.075 from left
to right~\cite{Laer}.  
   }
\end{figure}
~\\
\indent
In this section we have investigated the properties of the quark condensate
$\langle \bar{\psi} {\psi}\rangle$ alone at various values of the coupling.
However, here it is very difficult to immediately go over to a physical
temperature scale in the same way as in the previous section for the pure
gauge or gluon system. In what follows we shall look into the gluon condensate
in the presence of dynamical quarks. Here we know that the presence of the
quark masses are an immediate cause of scale symmetry breaking which of course
change the scale of the system. This in turn changes the beta function as well
as adds a term due to the mass renormalization. Thus the renormalization group
equations are changed accordingly. This effect we shall discuss more thoroughly
in the following section.



\section{Gluon Condensate from the Trace Anomaly in the Presence of Quarks
with finite Masses}

\noindent
The discovery of anomalous terms appearing as a finite value of the trace
of the energy momentum tensor was pointed out as a result of nonperturbative
evaluations in low-energy theorems~\cite{nonpert} many years ago. Furthermore,
it was also somewhat later realized how this factor arose with the process
of renormalization in quantum field theory which became known as the
trace anomaly~\cite{tracanom} since it was found in relation to an anomalous
trace of the energy momentum tensor.
~\\
\indent
     In the presence of massive quarks the trace of the energy-momentum tensor
takes the following form~\cite{tracanom} from the trace anomaly:   
\begin{equation}
\langle \Theta^{\mu}_{\mu} \rangle ~=~ m_q\langle \bar{\psi}_q{\psi}_q \rangle~
       +~\langle G^2 \rangle,
\label{eq:thetaquark} 
\end{equation}
\noindent
where $m_q$ is the light (renormalized) quark mass and ${\psi}_q$,
$\bar{\psi}_q$ represent the quark and antiquark fields respectively.
We include with these averages the renormalization group functions
$\beta(g)$ and $\gamma(g,m)$, which appear in this trace from the
renormalization process.  
~\\
\indent
     Now we would like to discuss the changes in the computational procedure 
which arise from the presence of dynamical quarks with a finite mass.  
There have been recently a number of computations of the thermodynamical 
quantities in full QCD with two flavors of staggered quarks~
\cite{BKT94,MILC96,MILC97}, and with four flavors~\cite{edwinqmfklat96,Eng5}.
These calculations are still not as accurate as those in pure gauge theory for
several reasons. The first is the prohibitive cost of obtaining statistics
similar to those obtained for pure QCD. So the error on the interaction 
measure is considerably larger. The second reason, perhaps more serious, lies 
in the effect of the quark masses currently simulated. They are still 
relatively heavy, which increases the contribution of the quark condensate 
term to the interaction measure. In fact, it is known that the vacuum
expectation values for heavy quarks~\cite{SVZ1}is proportional to the
vacuum gluon condensate or in the first approximation
\begin{equation} 
{\langle \bar{\psi}_q{\psi}_q \rangle}_{0}~=~
{-1 \over{12m_{q}}}{\langle G^2 \rangle_0}.
\label{eq:quarktogluecond}
\end{equation}
Furthermore, there is an additional 
difficulty in setting properly the temperature scale even to the extent of 
rather large changes in the critical temperature have been reported in the 
literature depending upon the method of extraction. For two flavors of quarks 
the values of \tc\/ lie between $0.140GeV$~\cite{MILC97} and about $0.170GeV$
~\cite{Edwin} which is considered presently a good estimate of the physical 
value for the critical temperature. 
~\\
\indent
     We now indicate briefly how the thermodynamical information is obtained 
for the equation of state in terms of the lattice quantities. We start with 
the expectation values of the lattice action $\langle S \rangle$, which now 
contains some improved contributions for the pure guage actions
~\cite{Eng5,EngKarSch} as well the contribution from the lattice fermions 
$\langle \bar{\chi}\chi \rangle$ as the chiral condensate as discussed 
in the last section. These quantities can be gotten from the partition 
function analogously to those in section 3. Explicitly it can be written as
\begin{equation}
\langle S \rangle~=~-{1\over{N^3_{\sigma}N_{\tau}}}~
{{\partial}\over{\partial{\beta}}}{\ln{\cal{Z}}},
\label{eq:actav}
\end{equation}
\noindent
and
\begin{equation}
\langle {\bar{\chi}\chi} \rangle~=~-{1\over{N^3_{\sigma}N_{\tau}}}~
{{\partial}\over{\partial{ma}}}{\ln{\cal{Z}}}.
\label{eq:ferav}
\end{equation}
\noindent
For the computation of the finite temperature analogous to the pure gauge
where the difference between $S_{0}$ and $S_{T}$ is used to compute the 
thermodynamics. Here we define
\begin{equation}
{\overline{\langle S \rangle}}~=~{\langle S \rangle}_0~-~{\langle S \rangle}_T
\label{eq:actavbar}
\end{equation}
\noindent
and
\begin{equation}
{\overline{\langle{\bar{\chi}\chi}\rangle}}~=~
{\langle {\bar{\chi}\chi}\rangle}_0~-~{\langle{\bar{\chi}\chi}\rangle}_T.
\label{eq:feravbar}
\end{equation}
\noindent
Now we define instead of the lattice beta function two similar quantities
called~\cite{Eng5}~$R_{\beta}$ and~$R_m$. We write
\begin{equation}
R_{\beta}~={d~{\beta}\over{d~{\ln}a}}, 
\label{eq:rbeta}
\end{equation}
\noindent
and
\begin{equation}
R_{m}~={d~{ma}\over{d~{\ln}a}}. 
\label{eq:rm}
\end{equation}
\noindent
We are now able to define a new interaction measure in the presence of
dynamical quarks $\Delta_{m}(T)$ in terms of these quantities, so that
\begin{equation}
  \Delta_{m}(T)~=~-{N^{4}_{\tau}}~{\left[R_{\beta}{\overline{\langle S\rangle}}
~-~R_m {\overline{\langle {\bar{\chi}\chi} \rangle}} \right]}.
\label{eq:rm}
\end{equation}
\noindent
Here we note the explicit effect of the quark condensate in the computation
of $\Delta_{m}(T)$.  
~\\
\indent
     Here it is appropriate to briefly explain another approach~\cite{Eng5}
to the setting of the temperature scale for the lattice computations. An 
effective coupling can be defined ${\beta}_{eff}$ in terms of the gluonic
part of the action as
\begin{equation}
{\beta}_{eff}~={12 \over{{\langle S \rangle}(\beta)}}. 
\label{eq:effcoup}
\end{equation}
\noindent
Then the dependence is used to calculate the derivative $R_{\beta}$ with
the help of the asymptotic two loop renormalization group equation. This
procedure~\cite{Eng5} also fixes the temperature scale so that
\begin{equation}
{T \over T_c}~=~{\left({\beta}_{eff} \over {\beta_c} \right)}^{-77/625}
     \exp{\left({{4\pi^2}\over 25}(\beta_{eff}-\beta_{c}) \right)}. 
\label{eq:tempscale}
\end{equation}
\noindent
This method allows us to establish the temperature in comparison to the 
critical temperature.
~\\
\indent
     We are now able to write down an equation for the 
temperature dependence of the thermally averaged trace of the energy momentum
tensor including the effects of the light quarks from $\Delta_{m}(T)$ so that
\begin{equation}
\langle \Theta^{\mu}_{\mu} \rangle_T~=~\Delta_{m}(T)\times T^4.
\label{eq:fulltheta}
\end{equation}
The thermally averaged gluon condensate is computed including the
light quarks in the trace anomaly using the equation~\eqref{eq:thetaquark}
and the interaction measure in $\langle \Theta^{\mu}_{\mu} \rangle_T$ to get
\begin{equation}
\langle G^2 \rangle_T ~=~ \langle G^2 \rangle_0
~+~m_q\langle\bar{\psi}_q{\psi}_q \rangle_0~
~-~m_q\langle\bar{\psi}_q{\psi}_q\rangle_T~
~-~\langle \Theta^{\mu}_{\mu}\rangle_T.
\label{eq:fullcond}
\end{equation}
It is possible to see from this equation that at very low temperatures 
the additional contribution to the temperature dependence of the gluon
condensate from the quark condensate is rather insignificant and disappears 
at zero temperature. However, in the range where the chiral symmetry is 
being restored there is an additional effect from the term
$\langle \bar{\psi}_q{\psi}_q\rangle_T$, which lowers $\langle G^2 \rangle_T $.
Well above \tc\/ after the chiral symmetry has been mostly restored the only 
remaining effect of the quark condensate is that of 
$m_q\langle\bar{\psi}_q{\psi}_q \rangle_0$. It is known~\cite{SVZ1} that
this term then the gluon condensate of the vacuum. Thus we expect~\cite{BoMi} 
that for the light quarks the temperature dependence can only be  
important below \tc\/. In the case of the chiral limit~$m_q \rightarrow 0$
the equation~\eqref{eq:fullcond} takes the form of Leutwyler's equation
~\eqref{eq:condef} as, of course, it should because Leutwyler used two 
massless quarks~\cite{Leut}. For the smaller values of the simulated quark 
masses in lattice units of 0.01 to 0.02 
$\langle \bar{\psi}_q{\psi}_q\rangle_T$ has mostly disappeared in the range 
where $\langle G^2 \rangle_T$ differs from $\langle G^2 \rangle_0$. 
~\\
\indent
     In order to see this effect, we look at Figure 3 
where the data of various simulations of finite temperature QCD with dynamical
quarks is used to show $\langle G^2 \rangle_T$ as a function of T. Included
in this figure is a plot of the pure gauge $SU(3)$ with its \tc\/ rescaled to
$0.150~GeV$, which is shown with the broken lines~\cite{BoMi}. The Bielefeld 
four flavor data~\cite{edwinqmfklat96,Eng5} is used to compute 
$\langle G^2 \rangle_T$~\cite{BoMi} shown in the open squares. The masses
in these simulations are rather large, 0.05 and 0.1 in lattice units. These
quite large masses do cause the quark terms to be more important at a 
considerably lower temperature. Also since there are four massive flavors,
the total contribution is larger. Thus at lower temperatures below \tc\/
the points for $\langle G^2 \rangle_T$ lie considerably below those values for 
the pure gauge theory. Starting at around $0.100~GeV$ we can see~\cite{BoMi}  
a difference from $\langle G^2 \rangle_0$. However, at high temperatures
we notice that the points are well above the pure gauge theory.
~\\
\indent
      For the two flavor quarks we show two sets of published data from the
MILC collaboration~\cite{BKT94,MILC96,MILC97}, both of which involved a 
rather extensive analysis of various thermodynamical quantities. The masses
in these cases are somewhat lighter, 0.0125 and 0.025 in lattice units. The 
earlier data~\cite{BKT94,MILC96},which has been used~\cite{BoMi} to compute 
$\langle G^2 \rangle_T$, is indicated by the solid squares in Figure 3.
Its behavior shows a smaller decrease than the four flavor data and, indeed,
follows more closely the pure $SU(3)$ data~\cite{Boyd} at lower temperatures.
At or slightly above \tc\/ it appears that the data points for
$\langle G^2 \rangle_T$ even rise~\cite{BoMi}. This apparent effect does
not seem to be very physically reasonnable or, at best, it is unexpected!
We would physically expect the raising of the system to higher temperatures
to continually reduce the amount of gluon condensate. However, the newer
data~\cite{MILC97}, which is indicated by the crosses in Figure 3, follows  
closely the earlier below \tc\/, but falls considerably below the solid
squares above the critical temperature. Nevertheless, the data points at
the higher temperatures still do not decrease in value very rapidly so that
within the errorbars the decondensation appears to remain flat. This
tendency is clearly very different from the pure gauge theory as shown in
Section 3.
\begin{figure}[tb]
  \begin{center}
    \leavevmode
     \epsfig{file=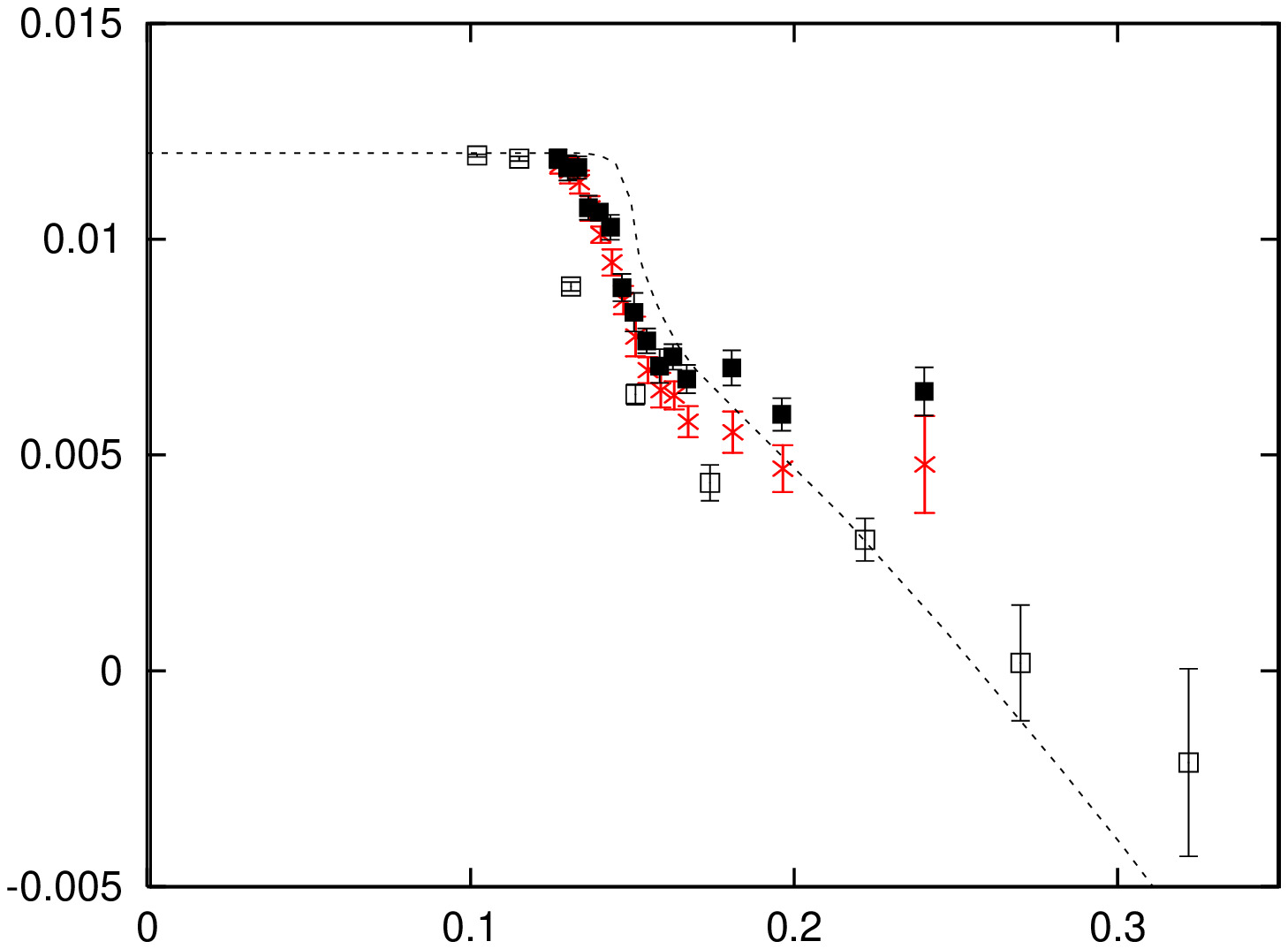,width=150mm,height=200mm}
     \caption{The gluon condensate $<G^{2}>_T$ in $GeV^4$ with dynamical quarks
      is plotted against the temperature T in $GeV$ in the following cases: 
      pure gauge theory~\cite{Boyd} (broken lines); 
      Bielefeld computation~\cite{Eng5} (open squares); 
      MILC'96~\cite{MILC96} (filled squares); MILC'97~\cite{MILC97} (crosses)} 
    \label{fig:full1}
  \end{center}
\end{figure}
~\\
\indent
As an end to this discussion  of the gluon condensate in QCD we will mention 
a few other points. Where in simulations on pure $SU(N_{c})$
gauge theory we could depend on considerable precision in the determination
of \tc\/ and $\Delta(T)$ as well as numerous other thermodynamical functions,
it is still not the case for the theory with dynamical quarks. The statistics 
for the numerical measurements are generally smaller. The determination of 
the temperature scale is thereby hindered so that it is harder to clearly 
specify a given quantity in terms of $T$. Thus, in general, we may state that
the accuracy for the full QCD is way down when compared to the computations  
of the pure lattice gauge theories. However, there is a point that arises
from the effect that the temperatures in full QCD are generally lower, so that
\DT\/ is much smaller~\cite{BoMi}. Here we can only speculate with the present
computations \cite{BKT94,MILC96,edwinqmfklat96,MILC97}. Nevertheless, there 
could be an indication of how the stability of the full QCD
keeps $\langle G^2 \rangle _T$ positive for $T~>~\tc$.
The condensates in full QCD have also been considered by Koch and
Brown~\cite{kochbrown}. However, the lattice data which they used were not
obtained using a non-perturbative method, nor was the temperature scale
obtained from the full non-perturbative beta-function. Finally we should
remark that the determination of the gluon condensate for the vacuum
is more significant in the presence of dynamical quarks. At the present
stage of the computations the newer MILC results\cite{MILC97} would prefer
the earlier value of $0.012~GeV^4$~\cite{SVZ1} in contrast to the newer value
of $0.0226~GeV^4$~\cite{Nari} only in the sense that the former favors a
larger proportion of the condensate to have vanished near the critical
temperature.  



\section{Anomalous Currents at Finite Temperature}
~\\
\noindent
In the previous sections we noticed that the fact that the trace
of the energy momentum tensor does not vanish for the strong interactions
has important implications for the equation of state. 
Here we shall discuss some more theoretical results relating to
$\Theta^{\mu}_{\mu}(T)$ or more exactly its relation to the corresponding
differential four-forms on a four-manifold coming from $G^{2}d^{4}x$, 
the gluon condensate in four dimensional space-time, where $d^{4}x$ is short
for the wedge product of the four different space-time differentials. 
This situation brings about certain properties with
respect to the dilatation current as well as the special conformal 
currents, both of which are not conserved. On the other hand we have the
anomalous chiral current which relates to the other four-form arising from~
${\bar{G}}Gd^{4}x$ resulting from the nonconservation of the chiral current.
In absolute magnitude this current has less importance at high temperatures for
disappearing quark masses since the chiral symmetry is then completely restored
. Nevertheless, it obviously plays a role near the deconfinement temperature
of a system with finite quark masses through the above noted changes in the
gluon condensate at finite temperature. 
~\\
\indent
     The dilatation current $D^{\mu}$ has been defined above in terms of 
the position four-vector $x^{\mu}$ and the energy momentum tensor
~$T^{\mu \nu}$ as simply the product $x_{\alpha}T^{\mu \alpha}$ 
as moments of the energy density. 
In the case of general energy momentum conservation one can find~\cite{Jack}
quite simply a relation to the equation of state.
We now look into a volume in four dimensional space-time ${\cal V}_{4}$
containing all the quarks and gluons at a fixed temperature $T$ in equilibrium.
The flow equation \eqref{eq:dilcurr} holds when the energy momentum and all
the (color) currents are conserved over the surface $\partial {\cal V}_{4}$
of the properly oriented four-volume ${\cal V}_{4}$, which yields
\begin{equation}
\oint_{\partial {\cal V}_{4}}{\cal D}_{\mu}dS^{\mu}~=
~\int_{{\cal V}_{4}} T^{\mu}_{\mu}dV_{4},
\label{eq:dyxleform}
\end{equation} 
We have already introduced~\cite{Mill}  the {\it dyxle} three-form as 
~${\cal D}_{\mu}dS^{\mu}$ on the three dimensional surface 
$\partial {\cal V}_{4}$. The dyxle is the dual form to 
~$D_{\mu}dx^{\mu}$ the dilatation current one-form~\cite{Flan}
in four dimensional space-time. It represents the flux through this closed 
surface acting as the boundry of the volume ${\cal V}_{4}$.
On the right hand side of \eqref{eq:dyxleform} the integrated form
~$\int_{{\cal V}_{4}} T^{\mu}_{\mu} dV_{4}$ is an action or flux 
integral involving the equation of state. Since $T^{\mu}_{\mu}>0$,
the action integral is not zero. This action integral gets quantized with the
fields through the renormalization process. It acts as the source term.
~\\
\indent
     An analogous form can be defined for the four special conformal currents
which we shall call the {\it fourspan}. The dual forms are derived from the 
equation~\eqref{eq:specconf} in a similar manner to the dyxle, which yields
\begin{equation}   
\oint_{\partial {\cal V}_{4}}{\cal K}^{\alpha}_{\mu}dS^{\mu}~=
~\int_{{\cal V}_{4}}{2{x^{\alpha}}}T^{\mu}_{\mu}dV_{4},
\label{eq:fourspanform}
\end{equation} 
~\\
\indent
Of most immediate interest to us here is really the finite temperature part 
of the dyxle~${\cal{D}}(T)_{\mu}{dS^{\mu}}$ in relation to the quark-gluon
condensates in QCD. This physical quantity represents the
flux as force through an area at a temperature $T$. Directly interpreted the
dyxle is the first moment in space-time of the energy-momentum. The vacuum
part just represents a fixed quantization where its value comes out of the
renormalization of the loops from the QCD beta and gamma functions. When we set
only ${\Theta}^{\mu}_{\mu}(T)$ into the dyxle equation~\eqref{eq:dyxleform}, 
we get the integral over a bounded region of space-time in terms of the
actual four-forms in vacuo and at finite temperature
$~\int_{{\cal V}_{4}}{\left(G^{2}_{0}~-~G^{2}(T)\right)}dV_{4}$, which
comes directly out of Leutwyler's equation~\eqref{eq:condef}. This integral
is clearly related to the interaction measure~$\Delta(T)$ in a bounded region
of space-time. Hereby the problem is reduced to one in homology relating to
the topology of the space-time. Unfortunately, The topological properties of 
the fourspan at finite temperature is not so directly related with the lattice
results. The four conformal currents are each related to higher moments of the
energy-momentum in space-time. Therefore, each of the four currents have a  
different moment structure relating to a different second moments 
of the energy-momentum. Here we are able at this time to say very little 
numerically about them.
~\\
\indent
The flux integrals arising in finite temperature QCD remind us of some
geometrically much simpler properties of the electromagnetic flux, which
is usually described in terms of Gauss' Law using a closed two dimensional 
surface through which the electromagnetic fields penetrate. The nature
of this two dimensional spatial surface is easily represented as a subspace
of a three dimensional Euclidean geometry. Our envisioning now becomes a bit
more difficult with a three dimensional surface for a four dimensional space-
time where we try to attibute this flux structure to a dual three-form with
the property that the dyxle has a "source term" arising from the equation of 
state in four dimensional space-time. This is the statement of the integral
form of the above equation~\eqref{eq:dyxleform}. It would be nice if we
were able to directly relate this statement to that of confinement or
deconfinement. However, as a phase transition it is presently unclear how
to relate these results to a single simple order parameter in all cases.



\section{Summary, Conclusions and Speculations}
\noindent
The main investigation of this paper involves the study of nonconserved 
anomalous currents using numerical results relating to the existence of
particular four-forms in finite temperature QCD. For the pure gauge theory
we consider only the trace anomaly and its implications on the gluon
condensate as discussed in the third section. The presence of quarks
with finite masses brings about an interplay between the renormalization 
causing scale breaking and the explicit scale breaking due to the masses.
Here both anomalies must be considered.
~\\
\indent
     Although the explicit nature of the two anomalies have quite different 
physical origins, the general effect on the gluon condensate at high
temperature is quite similar. The trace anomaly alone reflecting the breaking
of the scale and conformal symmetries, while the chiral anomaly comes from
the two different representations of $SL(2,\bf{C})$ representing the two
chiralities. In the case of the scale and conformal symmetries there is
never restoration, but only a change in the breaking due to the temperature
dependence. For the chiral symmetry it is resored at high temperatures up to 
the effects of the quark masses. For the larger quark masses the breaking
at high temperature remains considerable as we saw in the fourth section.
In the fifth section we looked at the combined effects of the two anomalies
but only for fairly small quark masses, that is well under the energy scale of
the critical temperature. Thus the explicit mass effect with the gluon
condensate arising with the quark condensate is very small, which lets us
use the numbers directly from the simulations. At the moment we can only
at most speculate about the role of the heavier quarks with a mass around the
energy value of \tc\/. Here a brief thought on it may be of value.
~\\
\indent
    The speculations are on a three quark model with two light quarks of less
than 10 MeV and one heavy quark of about 150 MeV representing u-, d- and 
s-quarks respectively. The energy scale of the s-quark's mass is very close to
that of \tc\/. Thus any sizable s-quark condensate would have a sizable 
contribution to the gluon condensate at finite temperature through equation 
(5.13). The very heavy quark condensates are given by 
\eqref{eq:quarktogluecond}, which gives an amount inversely proportional to 
the quark mass. Thus one would expect the charmed as well as, of course, the 
bottom and top quarks to have a very small role at \tc\/ in the decondensation
of the gluons. The very heavy quarks are those which provide the static gluon 
fields without the dynamical effects.



\section{Acknowledgements}

\medskip
The author would like to thank Rolf Baier and Krzysztof Redlich for many very 
helpful discussions.
Also a special thanks goes to Graham Boyd with whom many of the computations 
were carried out. He is grateful to the Bielefeld group for providing their 
data, and especially to J\"urgen Engels for the use of his programs and many 
valuable explanations of the lattice results. He would like to recognize the 
support of ZiF for Summer 1998 in the project "Multiscale Phenomena and their 
Simulation on Massively Parallel Computers" led by Frithjof Karsch and Helmut 
Satz. He would also like to thank Edwin Laermann for discussions and the use 
of the Bielefeld data for the chiral condensate and Andreas Peikert for
information on the two flavor critical temperature. Many thanks go to Carleton 
DeTar for sending personally the newer data from the MILC collaboration.
Finally a word of thanks goes to Toni Rebhan and all the organizers of the
"Workshop on Quantization, Generalized BRS Cohomology and Anomalies" 
at the Erwin Schr\"odinger Institute in Vienna, Austria.
~\\ 



\end{document}